\documentclass[conference]{IEEEtran}
\IEEEoverridecommandlockouts
\usepackage{cite}
\usepackage{amsmath,amssymb,amsfonts}
\usepackage{algorithmic}
\usepackage{graphicx}
\usepackage{textcomp}
\usepackage{xcolor}
\def\BibTeX{{\rm B\kern-.05em{\sc i\kern-.025em b}\kern-.08em
    T\kern-.1667em\lower.7ex\hbox{E}\kern-.125emX}}
\begin{document}

\title{Optimizing Quotient Filters using Graveyard Hashing\\
}

\author{\IEEEauthorblockN{Isabelle Quaye$^*$}
\IEEEauthorblockA{\textit{MIT CSAIL} \\
Cambridge, USA \\
iquaye@mit.edu}
\and
\IEEEauthorblockN{Temi Taylor$^*$}
\IEEEauthorblockA{\textit{MIT Game Lab} \\
Cambridge, USA \\
temit@mit.edu}
}

\maketitle

\begin{abstract}
We aim to improve the performance of the Quotient Filter at high load factors. Our Graveyard Filter is a variation of the Quotient Filter which incorporates Graveyard Hashing, a technique that uses tombstones to counteract the effects of primary clustering. We summarize our implementation of the graveyard filter and detail approaches to redistributing tombstones. Evaluating these variations under conditions similar to the original quotient filter paper, we found the performance of the graveyard filter to be competitive for insertion and query operations, with certain redistribution schemes showing stronger performance at high load factors. We discuss potential further improvements, such as using the current load factor to determine the employed redistribution approach.
\end{abstract}


\section{Introduction}
\textbf{Approximate Membership Query} data structures (\textbf{AMQs}) are key components of many network protocols and database systems. Given a set of inserted elements, they verify whether a queried element is present or absent (with high probability). \textbf{Quotient filters} are one common implementation, improving on the older bloom filter design by better exploiting data locality\cite{qf}. However, its reliance on linear probing leaves it vulnerable to \textbf{primary clustering}, causing its performance to degrade at high load factors. We introduce the \textbf{graveyard filter}, a variation of the quotient filter which incorporates the anti-clustering technique of \textbf{graveyard hashing}\cite{gh}. We compared several implementations of the graveyard filter to the quotient filter; in any given test, at least one version of the graveyard filter was competitive with the quotient filter at lower load factors, and at least one outperformed the quotient filter at high load factors. These findings suggest that some load-adaptive combination of our graveyard filter variants may consistently outperform the original quotient filter.

\section{Related Work}
Quotient filters\cite{qf} resemble hash tables that resolve collisions with linear probing. They are parameterized by a value $q$; inserted elements are indexed into the table by the first $q$ bits of their hashed value (their ``bucket''), and the values stored in the table are the remaining $r$ bits (such that the hash function returns values of size $q + r$). Entries in the table use additional bits to track metadata, associating values with the buckets they were meant to be inserted into. This allows queries to properly reconstruct the hashed values present in the table.

Graveyard hashing is based on the practice of implementing deletions in hash tables by replacing the deleted element with a \textbf{tombstone} marker. These are treated as empty spaces by insertions, reducing the need to shift around multiple elements. \cite{gh} observed that tombstones have the additional benefit of breaking apart the clusters formed by linear probing. They distribute tombstones evenly across the table at a frequency and density determined by the current load factor. By amortizing the costs of future insertions and preventing the formation of long clusters, this process reduces the expected time taken by insertions from quadratic to linear with respect to the load factor.

\section{Graveyard Filter}

\subsection{Tombstones and AMQ Operations}\label{TM}
Graveyard filters aim to utilize tombstones in a manner that does not disrupt the properties that quotient filters rely on for correctness. A new metadata bit in each table entry indicates whether or not the position contains a tombstone. For tombstones, the space that normally contains an inserted value instead stores the concatenated indices of its \textbf{predecessor} and \textbf{successor}. These refer to the runs to which the nearest non-tombstone elements in each direction belong.

Deletions swap elements to the end of their current run before converting them into tombstones, ensuring that future insertions won’t interrupt existing runs. Insertions that target tombstones use the predecessor and successor to ensure that a proper ordering of runs would be maintained; if not, they default to behaving similarly to ordinary quotient filters. Both of these operations update adjacent tombstones in the case of a new run being created or removed.

\subsection{Redistribution Policies}\label{AA}
Tombstones require periodic cleaning to avoid negatively impacting queries. Graveyard hashing entails inserting new tombstones at the same time as this cleaning -- a \textbf{redistribution} process which we must carefully time. We considered five different \textbf{redistribution policies}: 
\begin{enumerate}
    \item \textbf{No Redistribution}: We perform neither any form of cleanup nor any artificial insertion of tombstones.
    \item \textbf{Amortized Clean}: We only clean up tombstones encountered during a deletion or query. There is still no artificial insertion of tombstones.
    \item \textbf{Between-Runs}: We clean up tombstones within clusters, leaving one at the end of each run where possible. This redistribution is triggered whenever the number of insertions since the last cleanup exceeds $\frac{table\_size}{4x}$, where $x$ is $\frac{1}{1-load\_factor}$.
    \item \textbf{Clean-up}: This is similar to the Between-Runs policy, with the additional step of artificially inserting a new tombstone between each run.
    \item \textbf{Graveyard Hashing}: We insert $\frac{table\_size}{2x}$ tombstones evenly throughout the table, as in \cite{gh}. Because we keep tombstones at the end of runs, we shift these tombstones as necessary, which may merge runs or clusters.
\end{enumerate}
The Between-Runs, Clean-Up, and Graveyard Hashing policies invoke their redistribution operations after a certain number of insertions or deletions. This avoids penalizing queries if there are no insertions to utilize new tombstones.

\section{Results,  Analysis, and Conclusion}

We implemented the quotient filter and graveyard filter in C++\footnote{Code can be found here: https://github.com/dsert1/6.506-project}, including variations for each of our redistribution policies. Experiments were performed on a machine with an Intel Core i5-13500HX processor (2.50 GHz) and 32 GB of RAM, running Windows 11. During testing, filters were initialized with a hash function producing 32 bit values and $q=25$, yielding tables approximately 33.5 MB in size.

During experiments, we performed insertions or deletions until the table’s load factor changed by 5\%, followed by 30 seconds of random queries and 30 seconds of successful queries. Throughput was calculated by measuring the time taken (for insertions/deletions) or the number of operations completed (queries). Three execution patterns were evaluated: one only inserting elements, one only deleting, and one with a mixture of both (10\% insertions followed by 5\% deletions). The results of this third pattern are presented below.

\begin{center}
    \includegraphics[width=4cm]{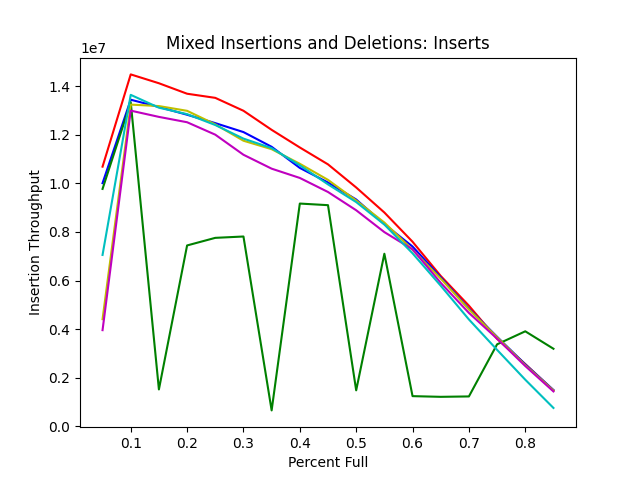}
    \includegraphics[width=4cm]{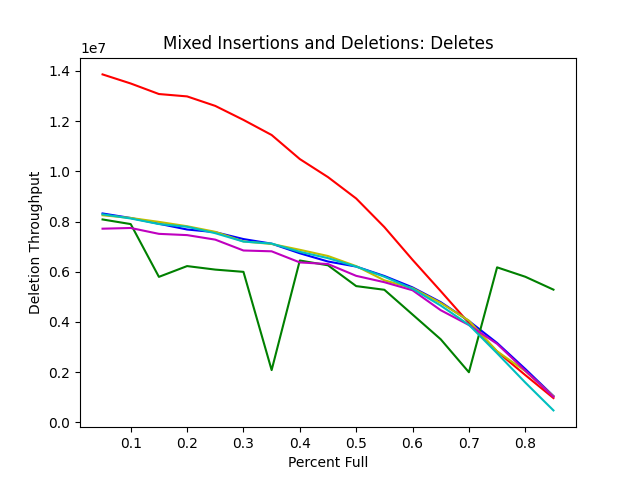}
\end{center}
\begin{center}
    \includegraphics[width=4cm]{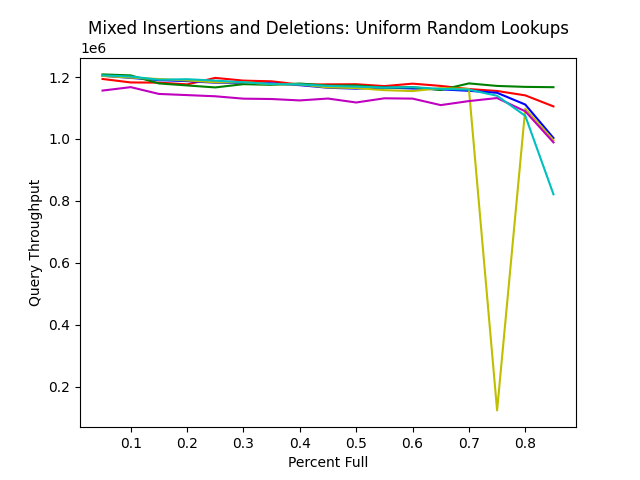}
    \includegraphics[width=4cm]{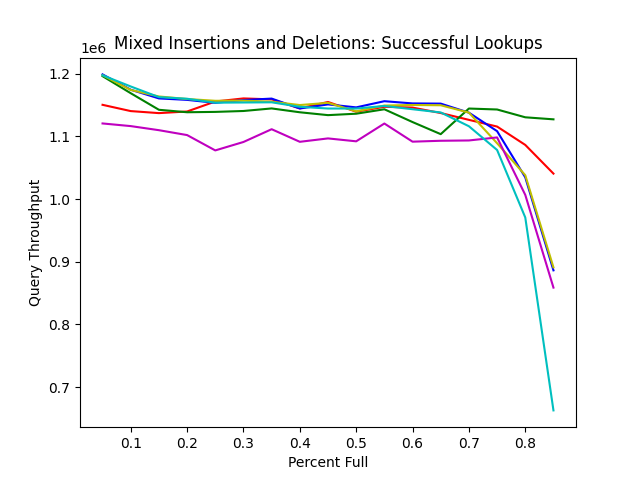}
\end{center}
\begin{center}
    \includegraphics[width=8cm]{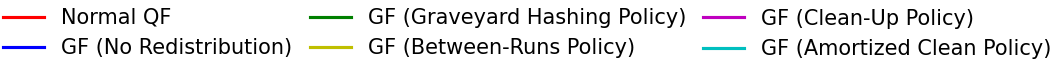}
\end{center}

The most prominent observation is that at least one variant of the graveyard filter starts to outperform the quotient filter at load factors of 75\% or higher. At lower load factors we see that the graveyard filter performs worse than the quotient filter for deletions, but for all other operations it remains competitive.

While the Graveyard Hashing policy most consistently outperformed the quotient filter at high load factors, it exhibited much less consistent behavior at lower load factors for insertions and deletions compared to other policies. This is likely because the movement of tombstones and merging of runs makes it more sensitive to the contents of the table and input stream than other policies.

As mentioned in \cite{qf}, the quotient filter's performance starts to suffer at higher load factors, falling below the bloom filter at 80\% fullness. The results here suggest that employing some graveyard filter policy when reaching this threshold could improve the overall performance. They also corroborate the claims made in \cite{gh} about tombstones breaking up clusters and improving the performance of insertions.

There is ample opportunity for future work, especially in terms of performance engineering. We list some possibilities below:

\begin{enumerate}
    \item \textbf{Parameter tuning}: The graveyard filter has a few parameters governing its performance, such as the frequency with which we clean up and redistribute tombstones. It is likely that an optimal frequency can be found for a given table size.
    \item \textbf{Auxiliary structures}: \cite{qf} suggests using auxiliary structures to store the starting index of buckets, eliminating the need to walk down long clusters to find where they start. It would be an interesting exercise to compare the impact this has on the performances of the quotient filter and the graveyard filter. In particular, we expect that the reduced policy maintenance overhead will further expose graveyard filter's insertion speed gains.
    \item \textbf{Policy analysis}: Further theoretical and empirical analysis of each redistribution policy and their behaviours under different workloads could reveal areas of improvement for each algorithm.
\end{enumerate}

\section*{Acknowledgment}
This project originated as an assignment for MIT's Algorithm Engineering course for Spring 2023 (6.506). We thank Deniz Sert for his contributions during the semester, as well as our instructors Charles Leiserson and Julian Shun for their advice and encouragement to submit this paper.


\begin{thebibliography}{00}
\bibitem{qf} M. A. Bender, M. Farach-Colton, R. Johnson, R. Kraner, B. C. Kuszmaul, D. Medjedovic, P. Montes, P. Shetty, R. P. Spillane, and E. Zadok. “Don’t thrash: how to cache your hash on flash.” Proc. VLDB Endow. 5, 11 (July 2012), 1627-1637. https://dl.acm.org/doi/10.14778/2350229.2350275
\bibitem{gh} M. A. Bender, B. C. Kuszmaul, and W. Kuszmaul. “Linear Probing Revisited: Tombstones Mark the Demise of Primary Clustering.” 2021 IEEE 62nd Annual Symposium on Foundations of Computer Science (FOCS), 1171-1182. https://arxiv.org/pdf/2107.01250.pdf
\end{thebibliography}
\end{document}